# Matrix Assisted Formation of Ferrihydrite Nanoparticles in a Siloxane/Poly(Oxyethylene) Nanohybrid

Nuno J. O. Silva[a], Vítor S. Amaral[a], Verónica de Zea Bermudez[b], Sílvia C. Nunes[b], Denis Ostrovskii[c], João Rocha[d] and Luís D. Carlos*[a]

[a] *Departmento de Física and CICECO, Universidade de Aveiro, 3810-193 Aveiro, Portugal Fax: 234 424965 Phonel: 234 370946; *E-mail: lcarlos@fis.ua.pt*
[b] *Department of Chemistry, University of Trás-os-Montes e Alto Douro, Quinta de Prados, Apartado 1013, 5001-911 Vila Real, Portugal.*
[c] *Department of Experimental Physics, Chalmers University of Technology 41296 Göteborg, Sweden.*
[d] *Departamento de Química and CICECO, Universidade de Aveiro, 3810-193 Aveiro, Portugal.*

Matrix-assisted formation of ferrihydrite, an iron oxide hydroxide analogue of the protein ferritin-core, in a sol-gel derived organic-inorganic hybrid is reported. The hybrid network (named di-ureasil) is composed of poly(oxyethylene) chains of different average polymer molecular weights grafted to siloxane domains by means of urea cross-linkages and accommodates ferrihydrite nanoparticles. Magnetic measurements, Fourier transform infrared and nuclear magnetic resonance spectroscopy reveal that the controlled modification of the polymer molecular weight allows the fine-tuning of the ability of the hybrid matrix to assist and promote iron coordination at the organic-inorganic interface and subsequent nucleation and growth of the ferrihydrite nanoparticles whose core size (2-4 nm) is tuned by the amount of iron incorporated. The polymer chain length, its arrangement and crystallinity, are key factors on the anchoring and formation of the ferrihydrite particles.

## Introduction

The development of self-assembled materials with the controlled formation of multifunctional entities for challenging applications in nano- and bio-technology is an active research theme. In this context, magnetic nanoparticles are of interest because they are widely used in magnetic storage[1] and have emergent applications in biomedicine for magnetic cell sorting[2] and magnetic fluid hyperthermia[3], among others. Most applications require controllable localized and selective formation of the nanoparticles and the possibility of grafting them with specific molecules. Recent approaches for the formation of magnetic nanoparticles involve the use of biological templates or matrix assisted assemblies. Examples of the former include the use of the ferritin protein coat to grow iron, manganese and uranium oxides[4,5] and a CoPt alloy for high-density data storage[6]. This concept was generalised to viral protein cages[7] and colloidal templates[8], providing complementary host-guest electronic interactions. Zeolites[9], polymers[10-12] and sol-gel derived silica glasses[13] have also been used to assist the formation of iron nanoparticles. The formation of the iron oxide hydroxide ferritin core in a sol-gel derived organosilica was recently reported[14]. The ferritin core holds up to 4000 iron atoms in a 5 nm radius ferrihydrite particle[15], exhibiting an interesting magnetic behaviour due to the surface canted spins and their coupling with the inner spins[16].

Here, we wish to report the matrix-assisted formation of ferrihydrite (iron oxide hydroxide) nanoparticles, analogues of the protein ferritin core, in a sol-gel derived organic-inorganic hybrid network (named di-ureasil). The latter consists of a siliceous backbone covalently grafted to poly(oxyethylene) (POE) chains of two distinct average molecular weight, $M_w$ (approximately 40.5 and 15.5 repeat units) by urea (NHC(=O)NH) cross-links (scheme 1)[17]. The polymer chains are a suitable and flexible medium to accommodate and wrap around iron-based particles of different sizes, preventing aggregation. This is reminiscent of the situation found in protein ferritin whose core is wrapped up in a protein coating. Moreover, the formation of ferrihydrite particles in the bio-inspired hybrids reported here is determined by the polymer chain length, its arrangement and crystallinity, thus pointing out the role of the matrix, whose cross-links are extensively involved in hydrogen bonding. This is a step forward relatively to the matrix-assisted assembly of ferritin core analogues in porous organosilica gels, in which the pore size (4-7 nm) determines the in situ assembly of ferrihydrite nanoparticles[14]. Further advantage of using organic-inorganic hybrids, such as di-ureasils, is the possibility of introducing different polymers and distinct organic-inorganic interface functionalities, as already been exploited for light emission applications[18-20].

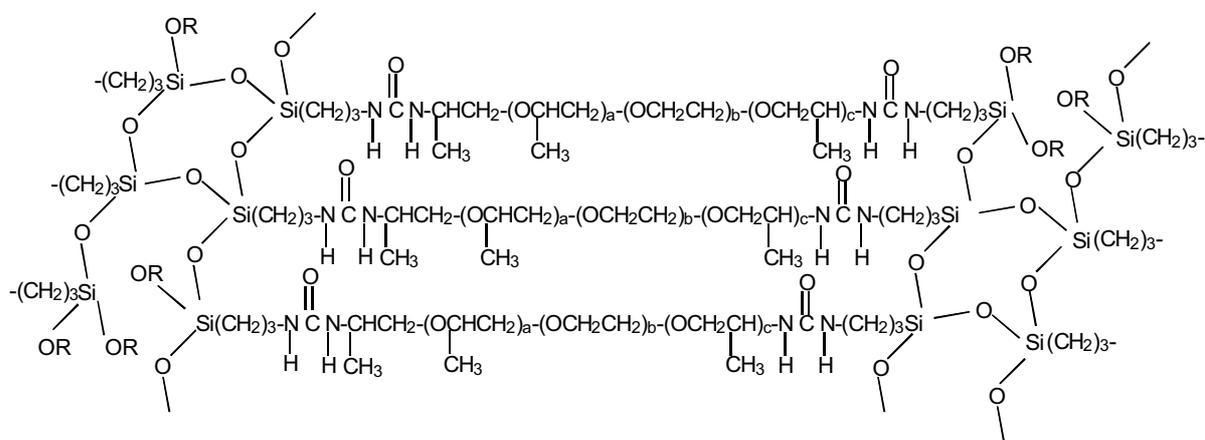

**Scheme 1** – Structure of di-ureasil hybrid, where a+c=2.5 and b=15.5 or 40.5 for, respectively, d-U(900) and d-U(2000) nanohybrids. R denotes hydroxyl or ethoxyl groups. The represented $T^1:T^2:T^3$ ratio corresponds approximately to $^{29}Si$ MAS NMR data for d-U(2000).

**Experimental**

**Sample preparation**

The synthesis of iron-doped di-ureasils has been described in detail elsewhere[21,23]. Two sets of samples with different POE chains length, termed d-U(2000) and d-U(900)[17], and iron concentrations up to 5.8 and 12.2 wt%, respectively, were prepared.

The preliminary step of the di-ureasils preparation involved the formation of urea linkages between the terminal NH groups of a doubly functional amine (α, ω-diamine poly(oxyethylene-*co*-oxypropylene)) available commercially as Jeffamine ED-2001® or Jeffamine ED-900® (Fluka, $M_w$=2000 and 900) and the isocyanate group of an alkoxysilane precursor (3-isocyanatepropyltriethoxysilane, ICPTES, Fluka) in tetrahydrofuran (THF, Merck) at room temperature (RT)[17]. A cross-linked hybrid precursor was, thus, obtained. The iron(III) nitrate nonahydrate (Fe(NO$_3$)$_3$.9H$_2$O, Aldrich) was incorporated in the second step of the synthetic procedure. In the case of the preparation of the d-U(2000)-based materials, an appropriate amount of this salt was dissolved in a mixture of ethanol and water (molar proportion ICPTES/CH$_3$CH$_2$OH/H$_2$O=1:4:1.5). This solution was added to the non-hydrolysed hybrid precursor prepared in the first stage. The resulting mixture was then stirred in a sealed flask for a few minutes at RT. The solution was cast into a mould. Gelation took place immediately. In the case of the synthesis of the iron-doped d-U(900)–based compounds, ethanol and water were added to the corresponding hybrid precursor, followed by the incorporation of the iron salt. The amount of ethanol added varied with salt concentration (5, 20 and 25 ml for the iron concentrations of 3.4, 9.5 and 12.2%, respectively). The mixture was stirred in a sealed flask for approximately 5-30 min, depending on the amount of salt added) at RT and then cast into a Teflon mould and left in a fume cupboard for 24 h. Gelation also occurred rapidly. In both cases, the mould was transferred to an oven at *ca*. 313 K for a period of 7 days. The sample was then aged for 3 weeks at *ca*. 353 K to form mechanically stable films. The structure of the non-doped di-ureasils was previously modelled as groups of siliceous domains with gyration radius of 0.5 nm correlated at an average distance $d_s$ of 4 nm and 3.2 nm, for d-U(2000) and d-U(900), respectively, embedded in a polymer-rich medium[22]. The polymer chains of the former hybrids are in a less folded and entangled conformation than those of d-U(900), as shown by the changes on the dependence between $d_s$ and $M_w$[22]. At the same time, the d-U(2000) polymer chains are essentially crystalline (trans-gauge-trans helicoidal conformation), unlike the amorphous polymer chains of the d-U(900) hybrids (trans-trans-trans ziz-zag conformation)[18]. Moreover, the glass transition temperature, $T_g$, of d-U(2000) is lower than the d-U(900) one ($T_g$=317 and 325 K, respectively), indicating that the polymer chains of the former are more flexible and the segmental chain motions are faster, involving a larger number of repeat units.

**Techniques**

Powder XRD: Measurements were performed at RT with a Philips X'Pert - MPD diffractometer using monochromated CuKα radiation (λ = 1.541 Å) in the 1.5 - 70º 2θ range at 0.05º resolution, and 35 s acquisition per step.

Magnetic Measurements: Data were collected on a SQUID (Superconducting Quantum Interference Device) magnetometer (model MPMS2, Quantum Design Inc) at IFIMUP - Universidade do Porto. The magnetic susceptibility was recorded at increasing temperatures under a field of 20 Oe, after an initial cooling from RT down to 4.5 K in the absence of the field (ZFC procedure) and in the presence of a field of 20 Oe (FC procedure).

FT-IR Spectroscopy: Mid-infrared spectra were acquired at RT using a Bruker 22 FT-IR spectrometer (model Vektor) placed inside a glove-box with dry argon gas atmosphere. The spectra were collected over the range 4000-370 cm-1 by averaging at least 150 scans at 2 cm$^{-1}$ spectral resolution of. The compounds (*ca*. 2 mg) were finely ground, mixed with approximately 175 mg of dried potassium bromide (Merck, spectroscopic grade) and pressed into pellets. To prevent the presence of water in the samples, the discs were dried at 90 ºC under vacuum (10$^{-6}$ mbar) for several days in the Buchi oven placed inside the same glove-box.

NMR Spectroscopy: $^{29}$Si magic-angle spinning (MAS) and $^{13}$C cross-polarization (CP) MAS NMR spectra were recorded on a Bruker Avance 400 (9.4 T) spectrometer at 79.49 and 100.62 MHz, respectively. $^{29}$Si MAS NMR spectra were recorded with 2 µs (equivalent to 30º) rf pulses, a recycle delay of 2 (d-U(900) Fe-containing samples) or 60 s and a 5.0 or 14.5 kHz (d-U(900) Fe-containing samples) spinning rate. $^{13}$C CP/MAS NMR spectra of d-U(2000) samples were recorded with a 4 µs $^1$H 90º pulse, 2 ms contact time, a recycle delay of 4 s and a spinning rate of 6-7 kHz. $^1$H high-power decoupled $^{13}$C MAS NMR spectra were recorded with 4 µs $^1$H 90º pulse, a 2 µs $^{13}$C 45º pulse and 80 s recycle delay. Chemical shifts are quoted in ppm from TMS.

**Results and discussion**

**XRD and SAXS**

Small-angle X-ray scattering (SAXS) studies on iron-doped d-U(2000) samples reveal the existence of iron-rich particles with gyration radius of 1.9, 3.0 and 3.7 nm, for iron concentrations of 1.2, 3.9 and 5.8 wt%, respectively[23], but no phase identification was performed at that point. Moreover, SAXS shows the absence of any particles or aggregates with other characteristic length in a scale up to *ca*. 30 nm. The mean interparticle distance is $d_s$=20 nm for all studied concentrations. The powder XRD patterns of the iron-doped d-U(2000) nanohybrids (Fig. 1, top diffractograms) show the existence of 6-line ferrihydrite particles and application of the Debye-Sherrer formula gives coherence lengths of the order of the values obtained with SAXS[23]. Thus, the iron-rich particles observed with SAXS can be assigned to ferrihydrite and the gyration radius there derived can be used as their mean size. In contrast, the powder XRD patterns of the iron-doped d-U(900) nanohybrids (Fig. 1, bottom diffractograms), indicate the absence of developed ferrihydrite particles even at iron concentrations 2 times higher than in the d-U(2000) system. The main feature of these diffractograms is a broad band centred at *ca.* 44° associated with the second order of diffraction of the siliceous domains. Small features at about 35° and 62° can be associated with the presence of minor amounts of Fe(III)-oxyhydroxy-nitrate phase, $FeO(OH)_{1-x}(NO_3)_x$ (0.2<x<0.3)[24]. The existence of ferrihydrite in 1.2% Fe-doped d-U(2000) and the 9.5% Fe-doped d-U(900) hybrids, which is not clearly established by the XRD patterns will be addressed in the next paragraphs.

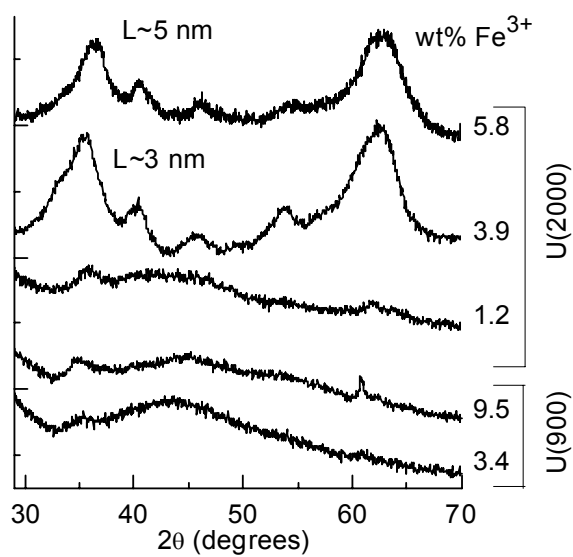

**Fig. 1** Powder XRD patterns of Fe$^{3+}$-doped d-U(2000) and d-U(900). The particles correlation length, L, was estimated applying the Debye-Sherrer law.

**Magnetic properties**

The temperature dependence of the magnetic susceptibility was recorded on heating after field-cooled (FC) and zero-field cooled (ZFC) procedures[21]. The ZFC curve of the 3.9% Fe-doped d-U(2000) sample exhibits a maximum at a mean blocking temperature $T_B$=13 K and a separation from the FC curve at about $T_F$=45 K (Fig. 2). The 1.2% Fe-doped d-U(2000) sample have similar $T_F$ temperature but no maximum can be discerned. In this case it is possible that the blocking temperature is below 5K or that the anisotropy barrier is so low that the field of measure (20 Oe) overrides it, as seen by Rao et al. [14]. In the ZFC curve, as the temperature is raised, the blocking spins are activated, producing an increase in the susceptibility. This effect is balanced with the thermal fluctuation of the unblocked spins, leading to the appearance of a maximum. We notice that in the most concentrated iron-doped di-ureasils, $T_B$ and the shape of FC curve below $T_B$ are similar to the case of ferritin [4,16]. In contrast, ferrihydrite aggregated nanopowders, with similar particle size but without any coating, have $T_B$~65 K and FC susceptibility curves essentially constant below $T_B$[25]. The magnetization of both the 1.2% and 3.9% Fe-doped d-U(2000) shows linear high field dependence and magnetic hysteresis below $T_F$, with the cycles centre shifted to negative fields. The characteristic coercive and exchange fields decrease rapidly with temperature, from values of the order of 1 kOe at 5 K to a few tens of Oe at $T_F$[21]. Although the XRD pattern of the 1.2% Fe-doped d-U(2000) is not conclusive about the existence of ferrihydrite, the magnetic features of this clearly establishes their existence. Moreover, from the magnetization data above $T_F$, the magnetic particle characteristic sizes were estimated [26] and are in agreement with the ferrihydrite gyration radius obtained by SAXS.

The magnetic properties the iron-doped d-U(900) are qualitatively different. The temperature dependence shows paramagnetic behaviour and can be fitted with a Curie law (Fig 2), which yields an average magnetic moment of 4.0 μB per iron atom. In addition, these nanohybrids have no thermal or field irreversibility, indicating that the majority of the iron ions do not interact magnetically. This confirms the absence of magnetic particles as ferrihydrite.

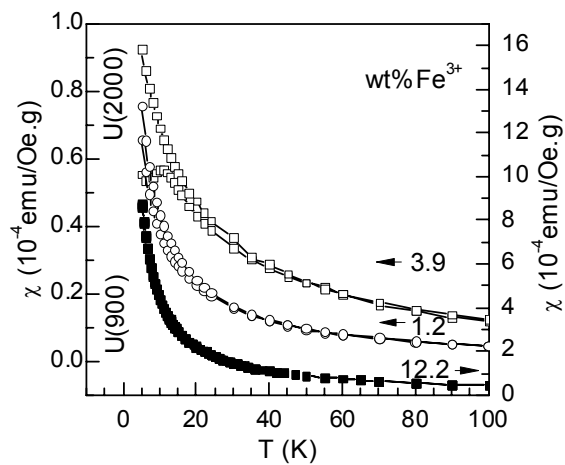

**Fig. 2** Temperature dependence of the magnetic susceptibility of d-U(2000) ($Fe^{3+}$ concentrations of 1.2 and 3.8 wt%) and d-U(900) di-ureasils ($Fe^{3+}$ concentration of 12.2 wt%), measured after field cooling (FC, upper curves in each set) and zero-field cooling (ZFC, lower curves in each set) procedures.

**FT-IR**

The "amide I" region of the FT-IR spectra of the iron-doped di-ureasils provides insight into the role played by the urea cross-links in the $Fe^{3+}$ coordination. The "amide I" mode is a complex vibration that receives a major contribution from the νC=O vibration[27]. As this is hydrogen bonding sensitive, the "amide I" envelope may be resolved into several components which correspond to different environments of C=O groups, known as hydrogen-bonded associations, aggregates or structures[27,28]. Fig. 3a shows that the profiles of the "amide I" envelope of the iron-doped d-U(2000) materials are dramatically different from that of the non-doped framework[17] and essentially independent of the $Fe^{3+}$ concentration. Upon introduction of the iron nitrate salt, a drastic decrease of the intensity of the component associated with "free" urea cross-links[19] (*ca.* 1750 $cm^{-1}$) and the concomitant increase of the intensity of the components produced by hydrogen-bonded POE/urea structures of low degree of disorder[17] (*ca.* 1719 $cm^{-1}$) occur. In contrast, the total spectral intensity of the features due to ordered POE/urea structures and self-assembled urea-urea aggregates[17] (*ca.* 1677 and 1647 $cm^{-1}$, respectively) seems to be almost unaffected by the presence of iron (Fig. 3a). These results indicate that a significant fraction of the available non-bonded urea carbonyl oxygen atoms of d-U(2000) interact with the $Fe^{3+}$ ions at the surface of the ferrihydrite nanoparticles. This means that the non-coordinated POE chains are massively requested to form hydrogen bonds of different strength with the N-H groups of the cross-links as a result of the formation of the ferrihydrite nanoparticles. Finally, it is evident from the FT-IR spectra of the doped samples in Fig. 3a that the relative proportion of carbonyl environments does not change with salt addition. This is consistent with the growth of the nanoparticles exclusively at the organic/inorganic interface and the build-up of a permanent disordered hydrogen bonded network throughout the di-ureasil host.

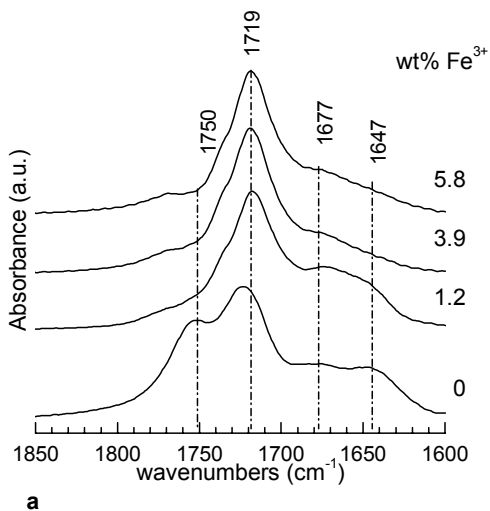
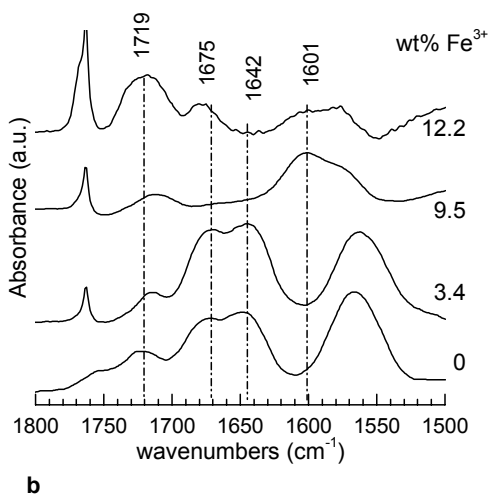
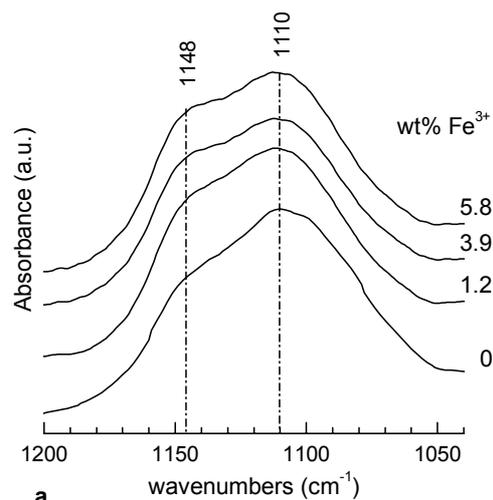
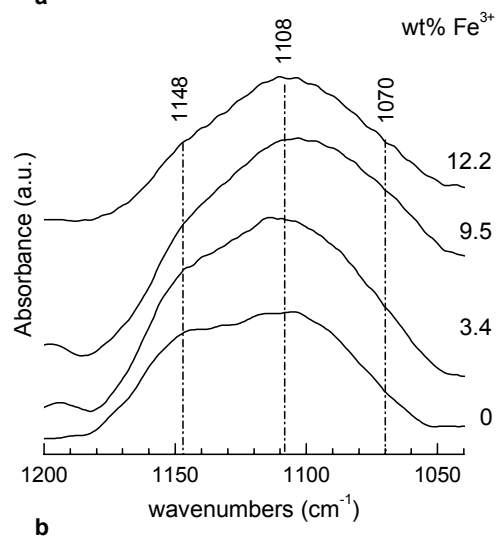

**Fig. 3** FT-IR spectra of the "amide I" envelope for undoped and $Fe^{3+}$-doped a) d-U(2000) and b) d-U(900) hybrids. Frequencies are indicated for the spectrum of the most concentrated hybrid.

**Fig. 4** FT-IR spectra in the vCOC region of undoped and $Fe^{3+}$-doped a) d-U(2000) and b) d-U(900) hybrids. Frequencies are indicated for the spectrum of the most concentrated hybrid.

The addition of iron nitrate to the d-U(900) host network strongly affects the "amide I" envelope (Fig. 3b). In contrast with the iron-doped d-U(2000), the changes produced are concentration dependent. At 3.4 wt% the component associated with "free" urea cross-links (*ca.* 1750 cm$^{-1}$) is no longer seen, proving that the "free" C=O groups are saturated by $Fe^{3+}$ ions. At 9.5 wt%, major modifications take place in this spectral range (Fig. 3b): although disordered hydrogen-bonded POE/urea structures are still present (*ca.* 1719 cm$^{-1}$), new associations (*ca.* 1601 cm$^{-1}$), considerably stronger than those initially present in d-U(900), form at the expense of a massive destruction of the ordered POE/urea and urea-urea aggregates (*ca.* 1675 and 1642 cm$^{-1}$, respectively). In the FT-IR spectrum of the most concentrated hybrid, the 1601 cm$^{-1}$ feature persists and some of the hydrogen-bonded aggregates initially present are formed again. These data show that the $Fe^{3+}$ coordination mechanism in the d-U(900) medium resembles globally that found in related POE/siloxane hybrid systems doped with other cations [29-31]. The interaction of the $Fe^{3+}$ ions with the polyether chains may be probed by FT-IR in the skeleton COC stretching (Fig 4). The FT-IR spectrum of the d-U(2000) matrix exhibits a broad band centred at *ca.* 1110 cm$^{-1}$ and a shoulder at *ca.* 1148 cm$^{-1}$ (Fig. 4a), ascribed to vCOC mode and to the coupled vibration of the vCOC and CH$_2$ rocking modes, respectively[17]. Because the intensity and frequency of both features remains practically unchanged in the presence of ferrihydrite nanoparticles, we conclude that the $Fe^{3+}$ ions do not coordinate to the ether oxygen atoms of the POE chains for the whole range of iron concentration examined. The situation is, however, different for the iron-doped d-U(900), as we note the growth of a shoulder in the low-frequency side of the vCOC envelope, around 1070 cm$^{-1}$ (Fig. 4b). This new feature is attributed to complexed polyether chains[30,31], i.e., to the vCOC vibration of oxyethylene units coordinated to $Fe^{3+}$ ions. Considering the relative intensity of the 1100 and 1070 cm$^{-1}$ bands, we conclude that at the highest concentration studied (12.2% wt) the proportion of non-complexed polymer chains still exceeds that of the complexed ones.

**NMR**

The $^{13}C$ CP/MAS NMR spectrum of the non-doped d-U(2000) is shown in Fig. 5a. The three peaks centred at about 10, 24 and 43 ppm are attributed to the three carbon atoms of -N(CH$_2$)$_3$Si-. The peaks observed near 18 and 70 ppm are ascribed to the few terminal propyl CH$_3$ groups and to the middle oxyethelene CH$_2$ groups of the organic chains, respectively (see Scheme 1). Finally, the resonance at *ca.* 159 ppm is assigned to the carbon atom of the C=O group of the urea linkage (Fig. 5a). Doping the di-ureasil host with iron results in a considerable broadening of all NMR resonances (Fig. 5a). For example, the full-width-at-half-maximum (FWHM) of the strongest peak, at *ca.* 70.5 ppm, increases fourfold from *ca.* 50 to 200 Hz. This broadening effect is even more important for d-U(900) samples, for which it was not possible to record $^{13}$C CP/MAS spectra. $^1$H high-power decoupled spectra with fast MAS (15 kHz) could be obtained but only for the sample with lower (3.4 wt%) Fe$^{3+}$ content (Fig. 5b), which exhibits a 70.5 ppm peak with a FWHM of *ca.* 470 Hz. This is due to the through-space interaction between the magnetic ferrihydrite particles and the $^{13}$C NMR nuclei. Thus, some polymeric chains are in close spatial proximity of the ferric nanoparticles or paramagnetic Fe ions in, respectively, d-U(2000) and d-U(900) samples. This effect is more pronounced in the latter case. Another important observation is that the signal-to-noise ratio of the d-U(2000) samples spectra decreases considerably as the iron concentration increases from 0 to 5.8 wt % (despite the fact that the number of accumulated transients was doubled), suggesting that NMR detects only a fraction of the total $^{13}$C present. Indeed, the resonances given by polymeric chains very close to the ferrihydrite nanoparticles are probably broadened beyond detection. Interestingly, the carbonyl resonance of the urea bridges is absent from the spectra of all iron-doped materials. It is not clear whether this is due to the coordination of carbonyl groups to iron atoms at the nanoparticles surface or, simply, because the carbonyl resonance broadens considerably upon doping and is buried in the baseline noise.

The $^{29}$Si MAS NMR spectra of the undoped d-U(2000) di-ureasil material (shown in Fig. 5c) displays three relatively sharp peaks in the -40 to -55 ppm region, attributed to RSi(O)(OH)$_2$ (T$^1$) surface groups. The resonances at *ca.* −59 and −68 ppm are assigned to RSi(O)$_2$(OH) (T$^2$) and R-Si(O)$_3$ (T$^3$) inner sites of the siliceous domains, respectively. Upon doping with iron the T$^1$ signals seem disappear from the spectra (not entirely clear because the signal-to-noise ratio of the spectra is poor). This is to be expected if the surface silicon environments are in close spatial proximity of the magnetic ferrihydrite nanoparticles. In addition (as observed in $^{13}$C NMR) the signal-to-noise ratio of the spectra decreases considerably as the materials iron concentration increases from 0 to 5.8 wt % (despite the fact that the number of accumulated transients was doubled), indicating that NMR detects only a fraction of the total $^{29}$Si present. It was only possible to record spectra of doped d-U(900) samples when fast (14.5 kHz) MAS was used (Fig 5d). The spectra of d-U(2000) and d-U(900) materials are similar, the main difference being that the $^{29}$Si NMR resonances in the latter are significantly broader.

## Conclusions

In these bio-inspired iron-doped di-ureasils the controlled modification of the average polymer molecular weight allows fine-tuning of the ability of the hybrid matrix to assist and promote iron coordination at the organic-inorganic interface (urea bridges) and subsequent nucleation and growth of nanoparticles (similar to protein ferritin cores). Di-ureasils with long organic segments successfully accommodated ferrihydrite nanoparticles (keeping constant density and increasing size, as the iron concentration increases). In contrast, in di-ureasils containing short organic segments, $Fe^{3+}$ ions are dispersed throughout the network and the formation of ferrihydrite nanoparticles does not occur. This difference is schematically depictured in Figure 6.

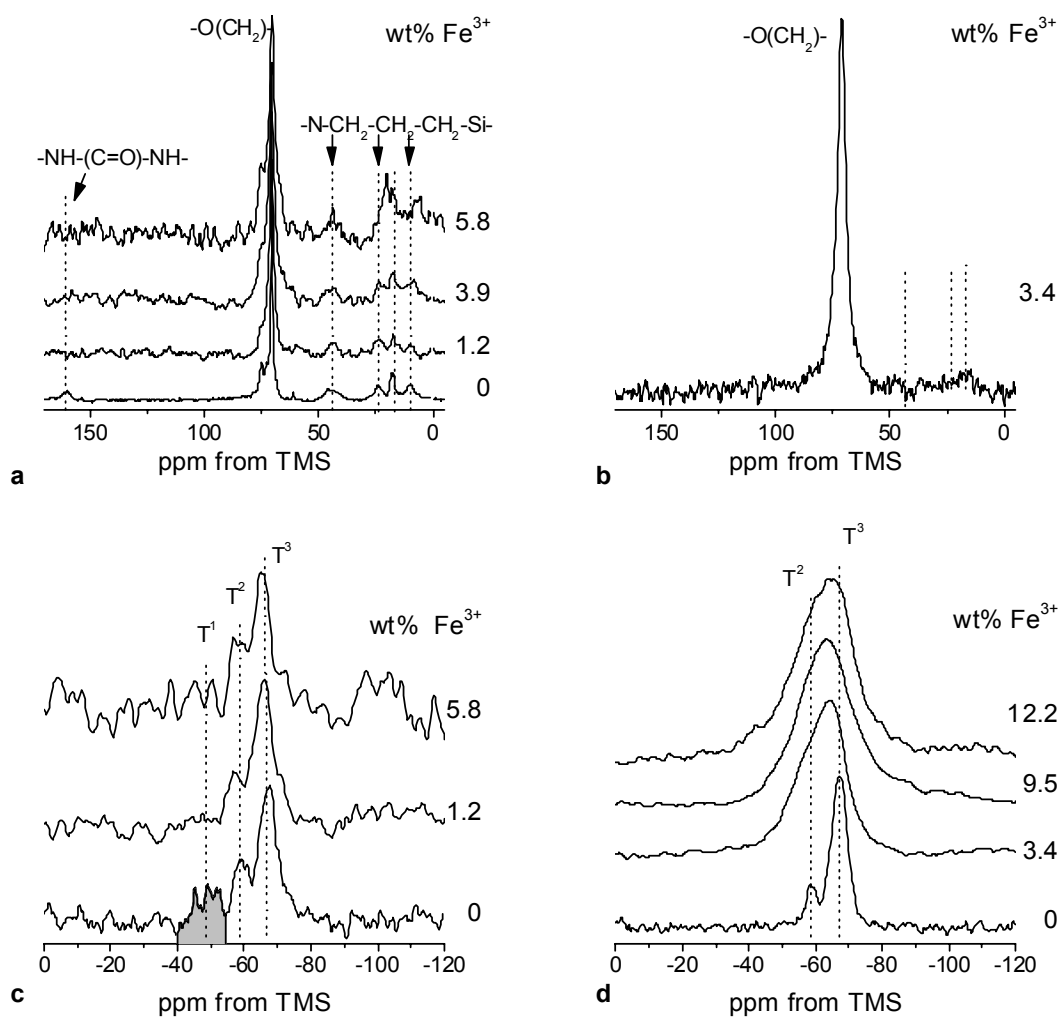

**Fig. 5.** $^{13}$C and $^{29}$Si NMR spectra for non-doped and the Fe$^{3+}$-doped di-ureasils. a) d-U(2000) $^{13}$C CP/MAS spectra (MAS 6-7 kHz); b) 3.4 wt% Fe-based d-U(900) $^1$H high-power decoupled $^{13}$C MAS spectrum (MAS 15 kHz); c) d-U(2000) $^{29}$Si MAS spectra (MAS 5 kHz); d) d-U(900) $^{29}$Si MAS (MAS 14.5 kHz) spectra.

Beyond the relevance of the distinct flexibility of the polymer chains in the two analogous hybrid hosts, from a local point of view all the spectroscopic data suggest that the significant differences detected are associated to the iron local coordination

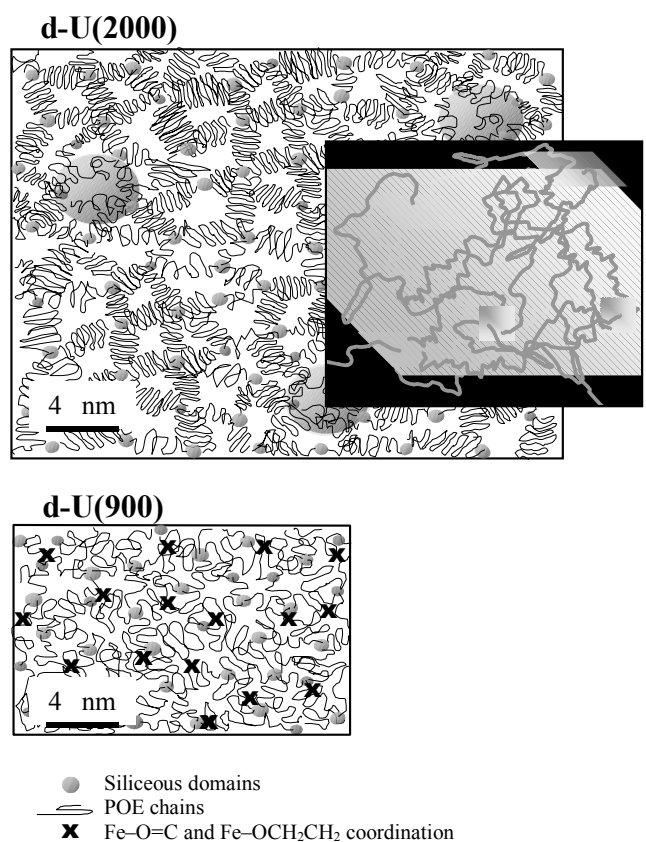

**d-U(2000)**

4 nm

**d-U(900)**

4 nm

- ● Siliceous domains
- ▱ POE chains
- ✗ Fe–O=C and Fe–OCH$_2$CH$_2$ coordination

**Fig. 6** Schematic representation of the structure of the Fe-doped d-U(2000) and d-U(900) hybrids. Small and large spheres depict the siloxane domains and the ferrihydrite nanoparticles, respectively (although not entirely correct, namely for the siliceous domains, this is the more simple approximation). The crosses represent the iron-carbonyl and the iron-oxyethylene coordination. The polymer chains are shown as coiled lines. The zoom in the d-U(2000) schematic representation illustrates the polymer chains wrapping around the ferritin core analogues.

and the magnitude and extension of the hydrogen-bonded urea-urea and urea-PEO associations. In d-U(2000) di-ureasils the iron atoms bond massively to the free carbonyl groups of the urea cross-links, at the organic-inorganic interface. Hence, the carbonyl-type oxygens are the preferential coordination sites, allowing the anchoring and subsequent nucleation of the ferrihydrite nanoparticles. This changes the conformations of the d-U(2000) polymer chains (presumably those located away from the nanoparticles) as these are massively requested to form hydrogen bonds with the N-H groups of the cross-links thus leading to the formation of a hydrogen-bonded network composed of POE/urea aggregates. The higher flexibility of the POE chains in the d-U(2000) di-ureasils and the faster segmental chain motions involving a larger number of repeat units allows higher iron diffusion rates in the host, resulting in efficient ferrihydrite particle growth. This flexibility enhances the ability of the polymer chains to wrap around the anchored ferrihydrite nanoparticles (Fig. 6).

The analogy between ferrihydrite nanoparticles incorporated in the long chain organic-inorganic di-ureasils and the protein-coated ferritin cores paves the way for the synthesis of novel bio-inspired hybrid materials with diverse polymer chain functionality. It also allows exploring the tuning of the iron solubilisation-precipitation processes.

## Acknowledgments

The financial support from FCT (POCTI/33653/CTM/00) and FEDER is gratefully acknowledged. NJOS and SCN acknowledge grants from FCT (SFRH/10383/2002 and SFRH/BD/13559/03).